\documentclass[aps,prl,twocolumn,showpacs,superscriptaddress,groupedaddress]{revtex4}
\usepackage{amssymb}
\usepackage{amsmath}
\usepackage{graphicx}
\usepackage{epstopdf}
\usepackage{subfigure}
\usepackage{natbib}
\usepackage{epsfig}
\usepackage{amsfonts}
\usepackage{mathrsfs}
\usepackage[toc,page,title,titletoc,header]{appendix}
\usepackage[colorlinks,linkcolor=blue,citecolor=blue,anchorcolor=blue]{hyperref}
\usepackage{dsfont,amsthm,amsbsy}

\begin{document}
\title{Entanglement Spectrum of Su-Schrieffer-Heeger-Hubbard Model}

\author{Bing-Tian Ye}
 \affiliation{School of Physics, Peking University, Beijing 100871, China}

\author{Liang-Zhu Mu}
\email{muliangzhu@pku.edu.cn}
 \affiliation{School of Physics, Peking University, Beijing 100871, China}

\author{Heng Fan}
\email{hfan@iphy.ac.cn}
 \affiliation{Institute of Physics,
Chinese Academy of Sciences, Beijing 100190, China}
\affiliation{Collaborative Innovation Center of Quantum Matter, Beijing 100190, China }

\date{\today}
\begin{abstract}
We investigate the entanglement spectrum of the ground state of Su-Schrieffer-Heeger-Hubbard model.
The topological phases of the model can be identified by degeneracy of the largest eigenvalues of entanglement spectrum.
The study of the periodic boundary condition is enough to obtain the phase diagram of the model, without
the consideration of the open boundary condition case.
Physical interpretation about the bulk-edge correspondence in the entanglement spectrum is presented.
The method of the entanglement spectrum can be applicable in studying other topological phases of matter.
\end{abstract}
\pacs{03.67.Mn, 03.65.Vf, 03.67.-a, 71.10.Fd}

\maketitle

\emph{Introduction.}---
The discovery of topological matters is an important advance of the well-developed band theory of condensed matter physics \cite{HasanKane,QiZhang}.
A great deal of interests focus, both theoretically and experimentally, on the topological
properties of those materials.
It is known that the topological properties of bulk state \cite{Berry,Nagaosa} and the existence of certain edge states existing in open boundary \cite{WuBernevig} can be seen as two symbols to distinguish different topological phases.
Also it is believed that there exists bulk-edge correspondence \cite{HasanKane,KaneMele}.
The bulk-edge correspondence can be demonstrated in the single-particle picture \cite{QiWu}.

Quantum entanglement is a key concept of quantum physics. It describes
quantum correlation between two subsystems, and acts as valuable resource
in quantum information processing.
Entanglement spectrum (ES) provides a full spectrum of the quantum entanglement by considering the local unitary
operations and Schmidt decomposition for a pure bipartite state.
Remarkably, it provides new perspective to investigate various physical characteristics,
first noticed in fractional quantum Hall effect states \cite{LiHaldane},  and in many other many-body systems.
The ES can be obtained by
finding the eigenvalues of reduced density matrix of the wave function for a system partitioned into two parts.
By considering that entanglement is a unique feature of quantum physics,
ES can illustrate the quantum properties of the system.
The method of ES has been successfully applied
to investigate complex paired superfluids \cite{Dubail}, edge states in Chern insulators \cite{Liu1,Bernevig,Liu2,Liu3},
spin-orbit coupled superconductors \cite{Borchmann}, etc.

In this Letter, we will use ES to investigate Su-Schrieffer-Heeger-Hubbard (SSHH) model which possesses topological phases.
The Su-Schrieffer-Heeger (SSH) model without interaction was introduced to describe polyacetylene \cite{SSH1,SSH2},
that can be shown to be a one-dimensional topological insulator \cite{Shenbook,GuoShen}.
Added with Hubbard interaction, the so-called SSHH model has even richer topological phases \cite{Wu}.
In the single-particle picture without Hubbard interaction, there are two topologically different phases \cite{GuoShen,Ryu}.
For periodic boundary condition (PBC), which means the chain is closed as a loop,
there is only bulk state (since there is no edge), the two distinct topological phases
can be characterized by Berry phase with 0 or $\pi$. For open boundary condition (OBC),
there exists edge states on the boundary. The entanglement entropy between two edges can be obtained by
subtracting the one half of the entanglement for PBC from the entanglement of OBC,
which shows a quantized behavior in two gapped phases \cite{Wu}.
Since ES can provide more information
than the single valued entanglement entropy, here in this Letter, we will study ES of this model.
For entanglement, naturally we need to divide the system into two blocks A and B, which
results in the changing of boundary condition from PBC to OBC for one subsystem where ES is defined.
Also the bulk-edge correspondence \cite{Shenbook} may warrant that ES for PBC
can provide concrete evidences in characterizing the topological phases.

\emph{The Model}---
The Hamiltonian of SSHH model is written as,
\begin{equation}
H_{SSHH}=H_{SSH}+H_{U}.
\end{equation}
The first term is SSH Hamiltonian with the following form,
\begin{equation}
\begin{split}
H_{SSH}=-[\sum^{L}_{i=odd,\sigma}&(t+\delta t)c^{\dag}_{i,\sigma} c^{}_{i+1,\sigma}\\
+&\sum^{L}_{i=even,\sigma}(t-\delta t)c^{\dag}_{i,\sigma} c^{}_{i+1,\sigma}]+H.c.,
\end{split}
\end{equation}
where $\sigma$ represents the spin, and $n_i=\sum_{\sigma}c^{\dag}_{i,\sigma} c^{}_{i+1,\sigma}$. For simplicity, we choose $t=1$ in this paper. The last term is the well-known Hubbard interaction $H_{U}=\frac{U}{2}\sum(n_i-1)^2$ originating from fermions with different spins on the same site. As usual, we will only pay attention to half-filled case in this paper \cite{SSH2,Shenbook}. In this condition, the Hubbard interaction is equivalent to
\begin{equation}
H_{U}=U\sum^L_{i=1}n_{i\uparrow}n_{i\downarrow}.
\end{equation}

Considering the convention that two neighboring sites $i$ (odd) and $i+1$ (even) are combined to be seen as one unit cell \cite{Shenbook}, the original closed chain and the cutted subsystem should have an even number of sites. Furthermore, we cut one end of the subsystem at the first site of the closed chain, so that the subsystem has exactly the same Hamiltonian as the original closed chain without ambiguity of site number,
but with different boundary condition (OBC).

\emph{The free-particle case}---
When without $H_{U}$, the ground state of the whole system can be derived from the single particle picture. The density matrix is in the following form \cite{Cheong},
\begin{equation}
\rho=\mathrm{det}(I-G)\mathrm{exp}(\sum_{ij}[\mathrm{ln}G(I-G)^{-1}]_{ij}c^{\dag}_i c^{}_j),
\end{equation}
with the Green function matrix $G_{ij}=\langle c^{\dag}_i c^{}_j\rangle$, where $c^{\dag}_i$ and $c^{}_j$ are fermion creation and annihilation operators acting on sites $i$ and $j$. It has been proved that the reduced density matrix in free Fermions case is of similar form, which means the reduced density matrix of subsystem A can be written as \cite{Cheong},
\begin{equation}
\rho_A=\mathrm{det}(I-G)\mathrm{exp}(\sum_{i,j\in A}[\mathrm{ln}G(I-G)^{-1}]_{ij}c^{\dag}_i c^{}_j),
\end{equation}
which means sites $i$ and $j$ only belong to the subsystem A. By diagonalizing the Green function matrix $G_{ij}$, we can derive,
\begin{equation}
\rho_A=\mathrm{exp} \{\sum_k\mathrm{ln}(1-f_k) +\sum_{l} [\mathrm{ln}f_l(1-f_l)^{-1}]d^\dag_l d_l\},
\end{equation}
in which $d^\dag_l$ is a new set of creation operators that can be obtained from $c^\dag_i$ by a unitary transformation and $f_{l}$ is the corresponding eigenvalue of $G_{ij}$. With the newly defined single particle states, the reduced density matrix of A is also diagonalized. We define $U_0$ to be those single particle states with filling particles, and $U_1$ to be those without filling particles.
The entanglement spectrum, which can be obtained as minus logarithm of the eigenvalues of the reduced density operator,
can be listed as the following,
\begin{equation}
    \epsilon_i=\mathrm{ln} [\prod_{k \in U_0}(1-f_k) \prod_{l \in U_1} f_l].
\end{equation}

\begin{figure}[h]
\begin{minipage}{0.45\linewidth}
\small{(a)} $\delta t=0.2$
\includegraphics[width=0.95\linewidth]{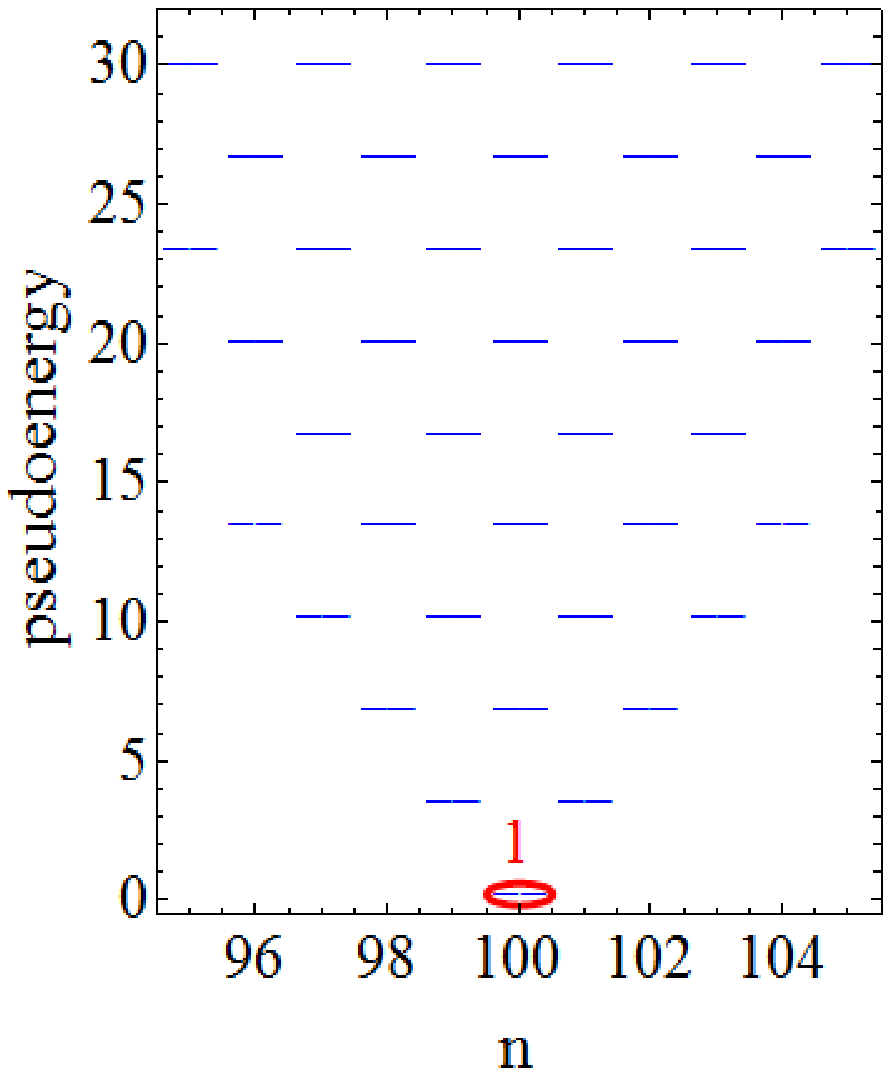}
\centering
\end{minipage}
\begin{minipage}{0.45\linewidth}
\small{(b)} $\delta t=0.4$
\includegraphics[width=0.95\linewidth]{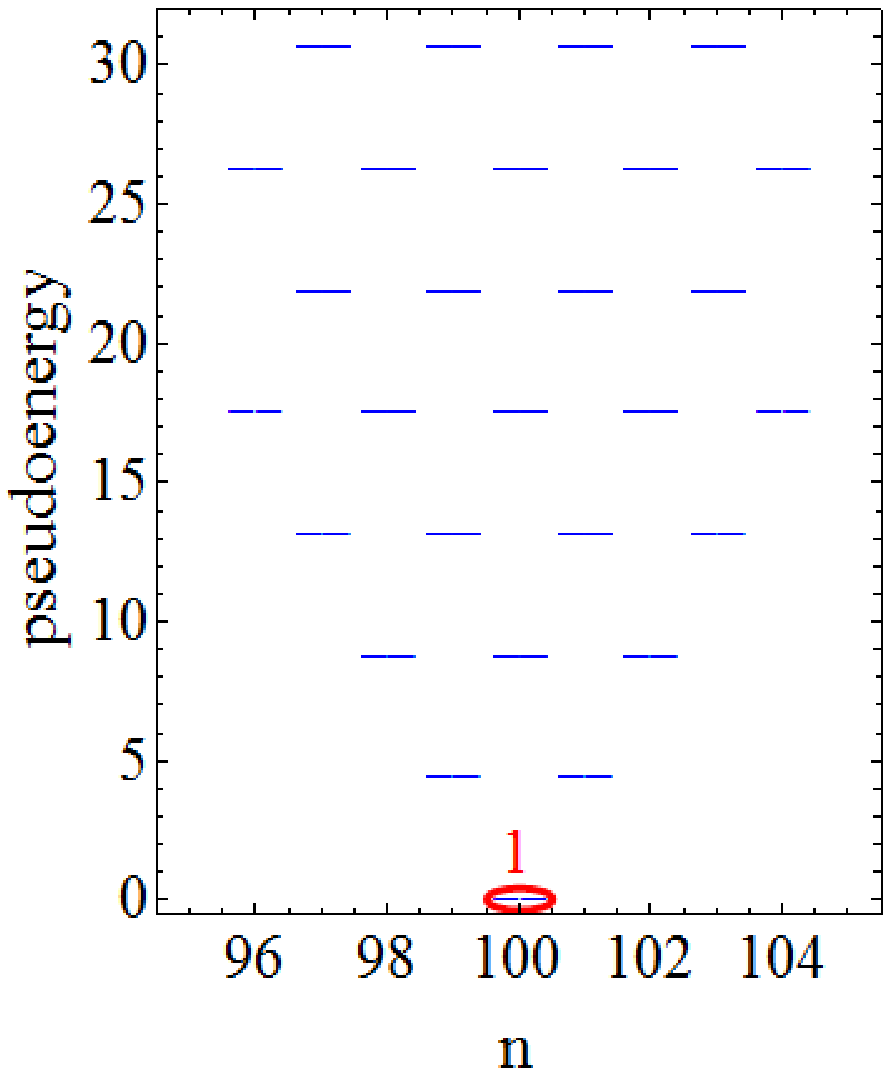}
\centering
\end{minipage}
\begin{minipage}{0.45\linewidth}
\small{(c)} $\delta t=-0.2$
\includegraphics[width=0.95\linewidth]{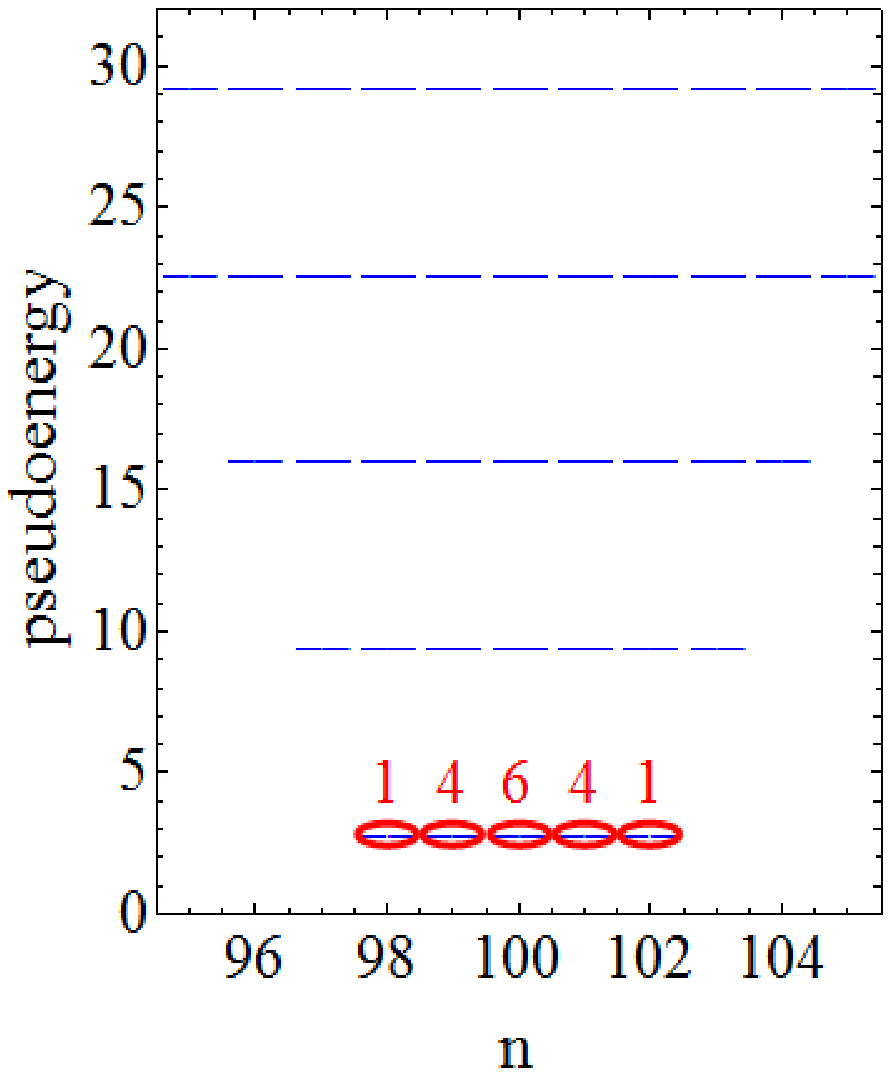}
\centering
\end{minipage}
\begin{minipage}{0.45\linewidth}
\small{(d)} $\delta t=-0.4$
\includegraphics[width=0.95\linewidth]{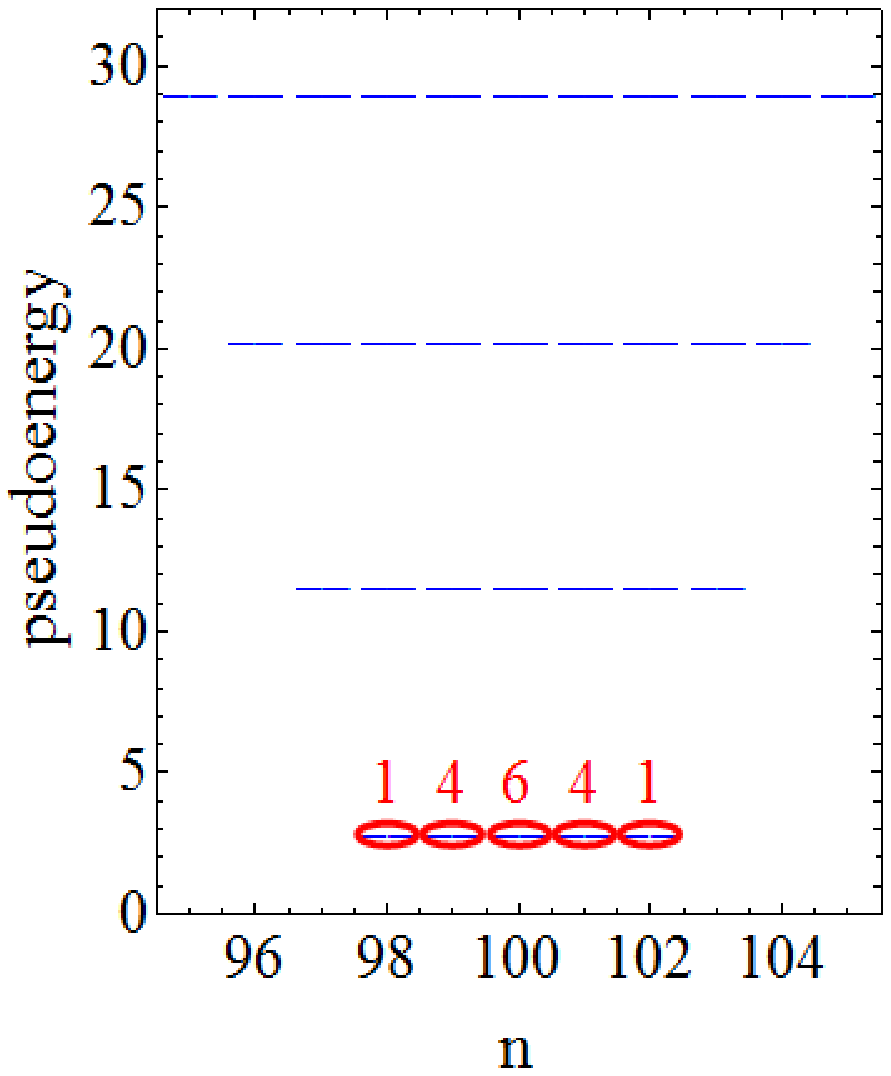}
\centering
\end{minipage}
\caption{\label{freespectrum} The chain length L is 200, and the subsystem is cut off with the length of 100. The spectrum are plotted with different $\delta t$. The number above each ground spectral line is the degree of degeneracy, so the total degree of degeneracy in the ground spectrum is the sum of these numbers.}
\end{figure}

Since the particle number is a good quantum number in the system, we can count the number of particles remained in the subsystem of each component corresponding to each spectral line. Additionally, we will show later that the particle number can be used to distinguish two slightly different phases when there is interaction. Considering all above, we plot the spectrum with the remaining particle number to be the horizontal ordinate. Figure~\ref{freespectrum} shows clearly the difference in the entanglement spectrum of different topological phases: in $\delta t>0$ condition, the ground spectral line is non-degenerate, while in $\delta t<0$ condition, the ground spectral lines are 16-fold degenerate. We claim that the exact value of $\delta t$ does not affect the degree of degeneracy, which means the degeneracy is a signature (symbol) of different phases.
We remark that ground spectral line corresponds to the largest eigenvalues of the reduced density operator in our partition.

\emph{Interacting-particle case.}---
We nest investigate the robustness of the topological ordered phase with interaction as disturbance,
where the system has Hubbard interaction $H_{U}$ in Hamiltonian.
We use Arnoldi method, which is an effective algorithm in finding the largest eigenvalues \cite{Arnoldi,Arnoldi1}, to achieve the ground state of the whole system and the reduced density matrix of subsystem A. Since the particle number is a good quantum number, the reduced density matrix is block diagonal, so we just need to diagonalize each block to find the entanglement spectrum.\\
\begin{figure}[h]
\begin{minipage}{0.45\linewidth}
\small{(a)} $\delta t=0.4,U=3$
\includegraphics[width=0.95\linewidth]{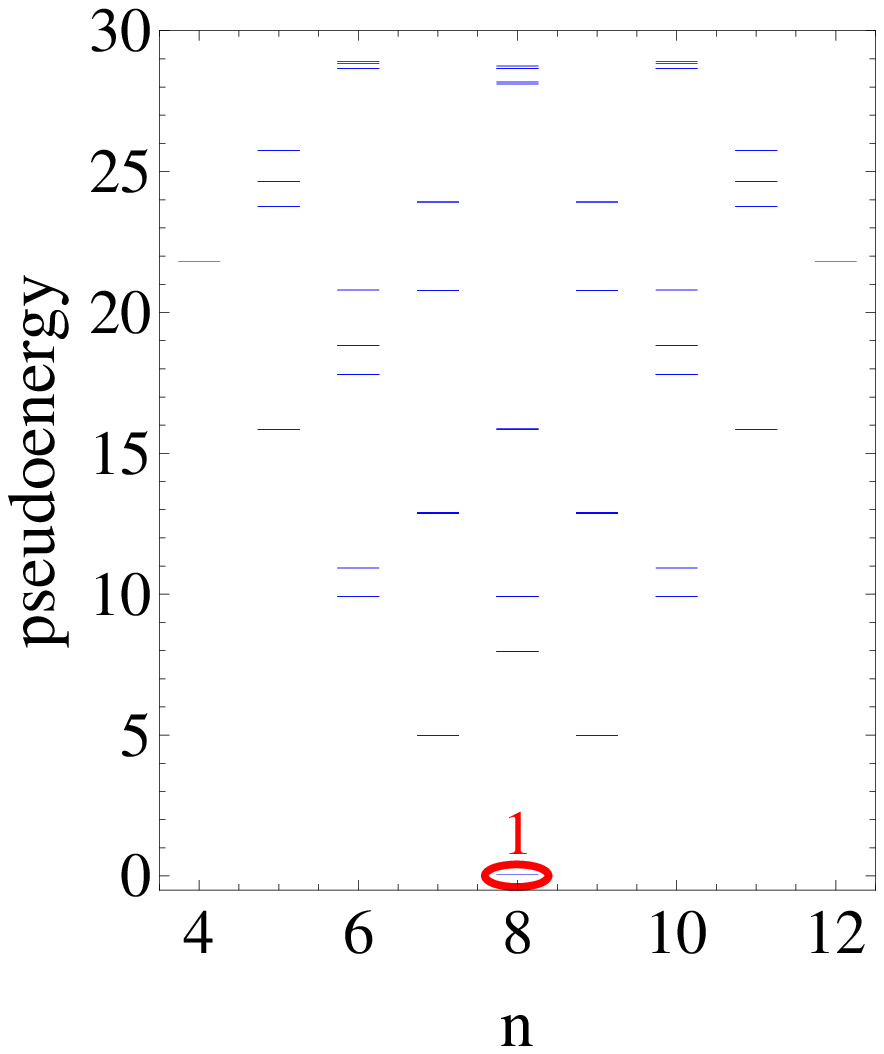}
\centering
\end{minipage}
\begin{minipage}{0.45\linewidth}
\small{(b)} $\delta t=0.4,U=-3$
\includegraphics[width=0.95\linewidth]{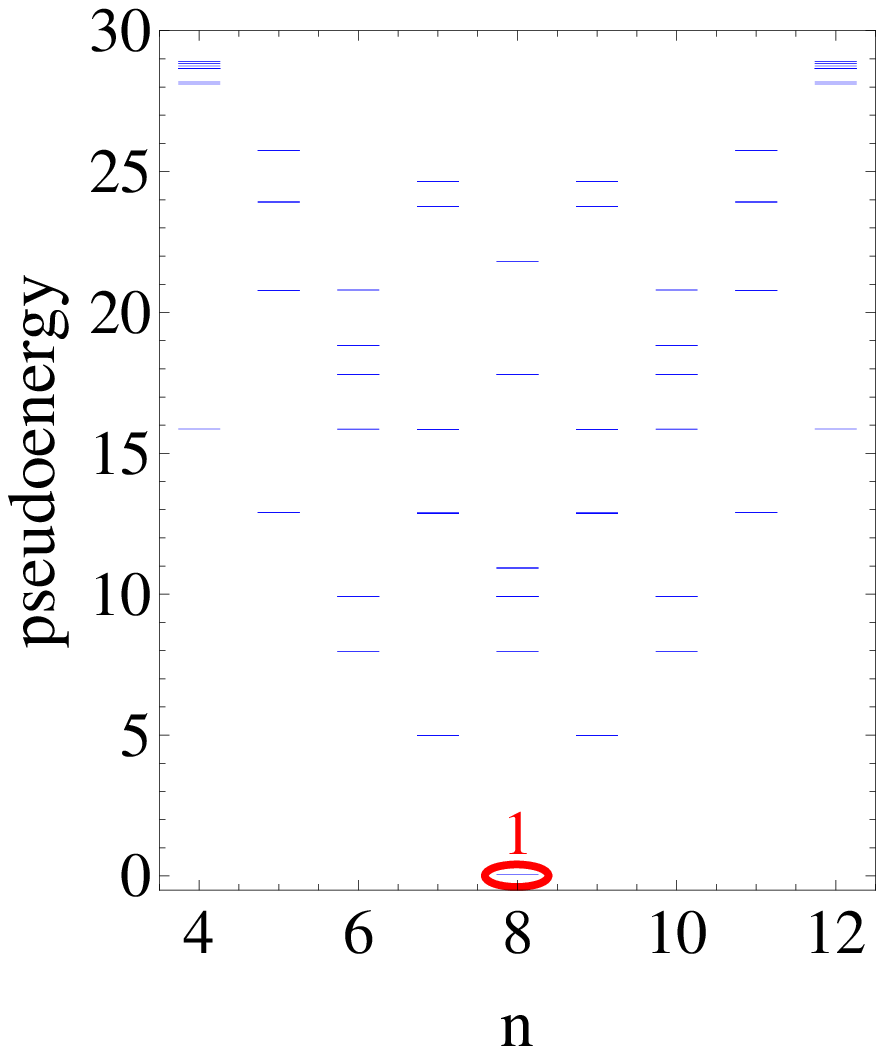}
\centering
\end{minipage}
\begin{minipage}{0.45\linewidth}
\small{(c)} $\delta t=-0.4,U=3$
\includegraphics[width=0.95\linewidth]{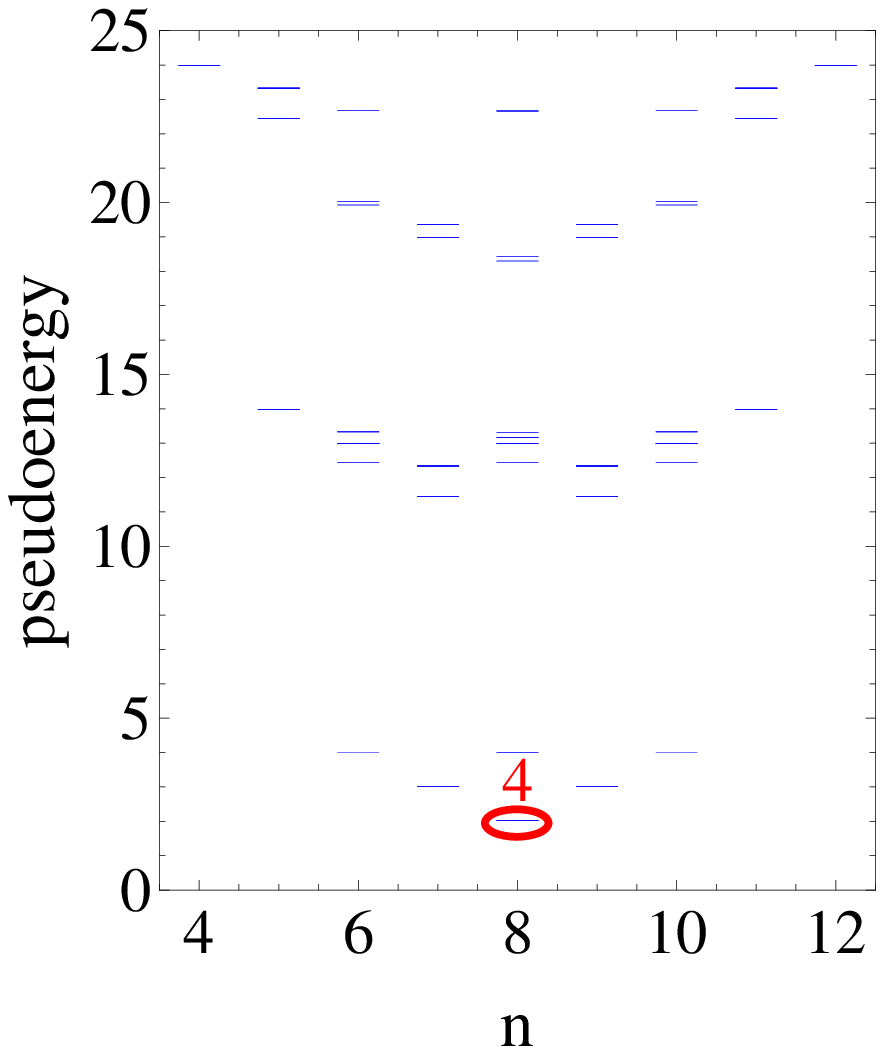}
\centering
\end{minipage}
\begin{minipage}{0.45\linewidth}
\small{(d)} $\delta t=-0.4,U=-3$
\includegraphics[width=0.95\linewidth]{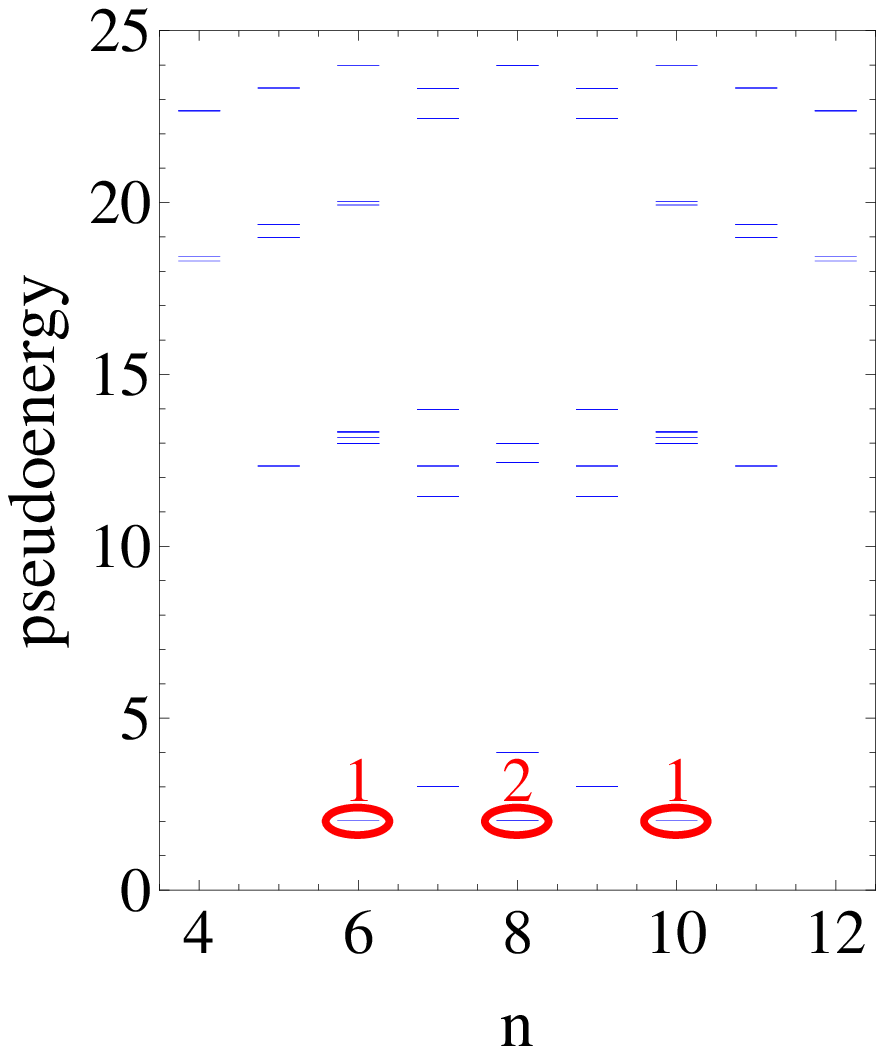}
\centering
\end{minipage}
\begin{minipage}{0.45\linewidth}
\small{(e)} $\delta t=0.3,U=5$
\includegraphics[width=0.95\linewidth]{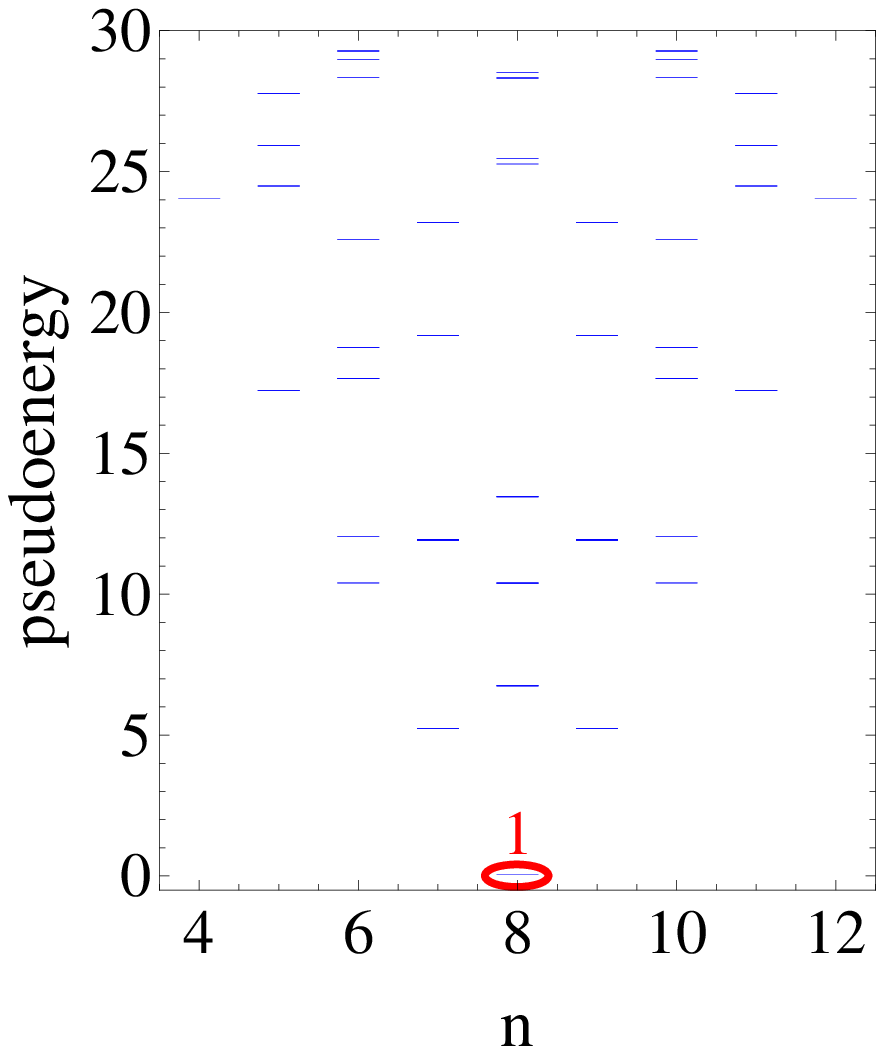}
\centering
\end{minipage}
\begin{minipage}{0.45\linewidth}
\small{(f)} $\delta t=-0.3,U=-5$
\includegraphics[width=0.95\linewidth]{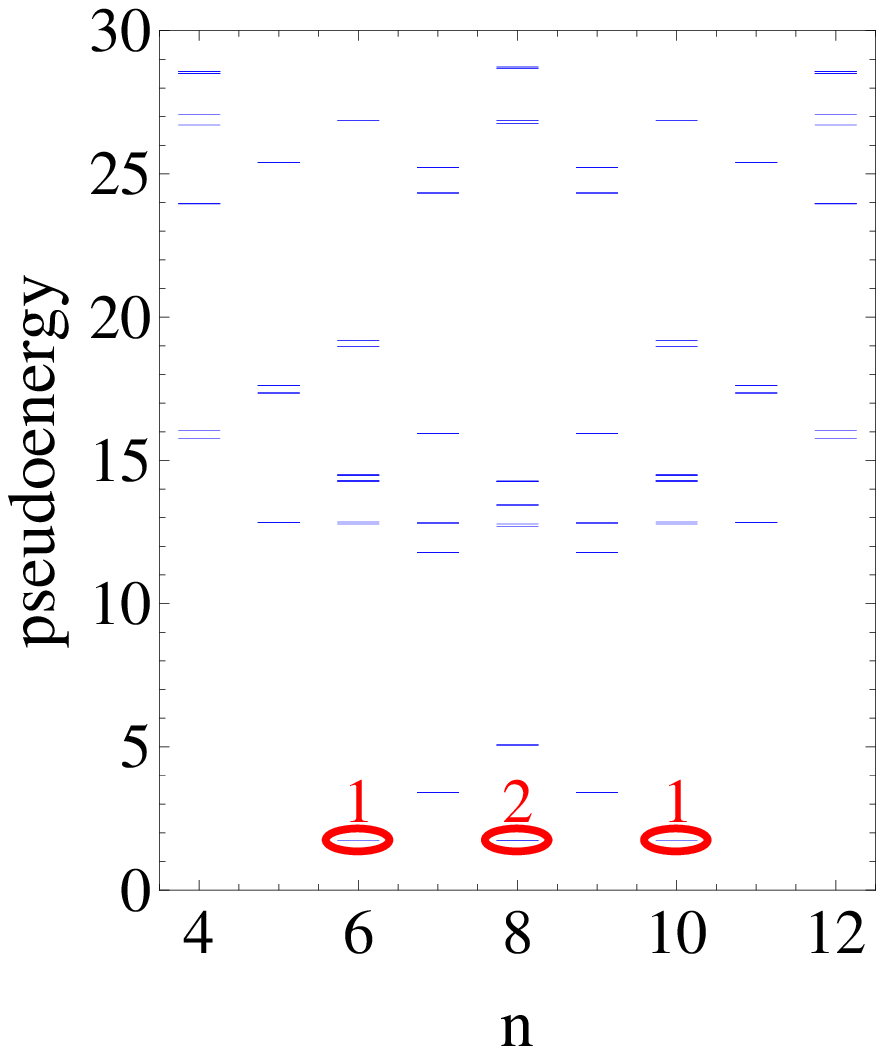}
\centering
\end{minipage}\\
\caption{\label{interactingspectrum} The chain length $L$ is 16. The spectrum are plotted with different combinations of $\delta t$ and $U$. The number above each ground spectral line is also the degree of degeneracy.}
\end{figure}
As shown in Fig.~\ref{interactingspectrum}, when there is interaction, in $\delta t<0$ condition, the ground spectral lines are 4-fold degeneracy, while in the $\delta t>0$ case, it is still non-degenerate. The results are slightly different in $U>0$ and $U<0$ conditions, i.e., the distribution of the degree of degeneracy according to particle number is different. Also, the exact values of $\delta t$ and $U$ do not affect the degeneracy of the spectral lines.

\begin{figure}[h]
\centering
\includegraphics[width=0.8\linewidth]{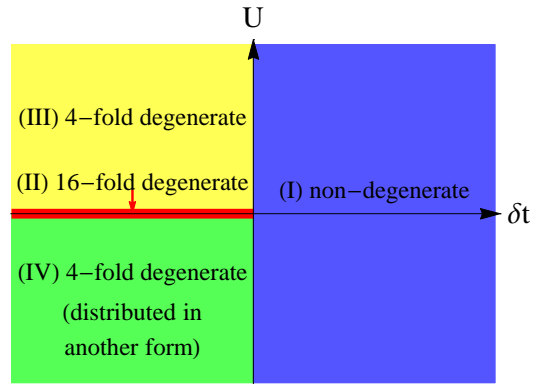}\\
\caption{\label{phasediagram} Different colors (gray levels), also labeled as I, II, III and IV, represent different phases according to the degeneracy of the ground entanglement spectrum.}
\end{figure}

According to the degeneracy of ES in different regimes of the model, we can draw the phase diagram shown in Fig.~\ref{phasediagram}. The upper half of the phase diagram can also be derived by investigate the entanglement entropy \cite{Wu}. In that method, the entanglement entropy is defined as the
entropy difference between PBC and OBC, which means that both entanglement entropies of PBC and OBC are needed. However in our method, the spectrum can be acquired from only PBC. Furthermore, by distinguishing the different distributions of degenerate states according to their remaining particle number in the two 4-fold degeneracy conditions, we can determine two slightly different phases, phase III and phase IV in Fig.~\ref{phasediagram},
they cannot be distinguished by only investigating entanglement entropy.

\begin{figure}[h]
\begin{minipage}{0.45\linewidth}
\small{(a)} $\delta t=0.4,U=3$
\includegraphics[width=0.95\linewidth]{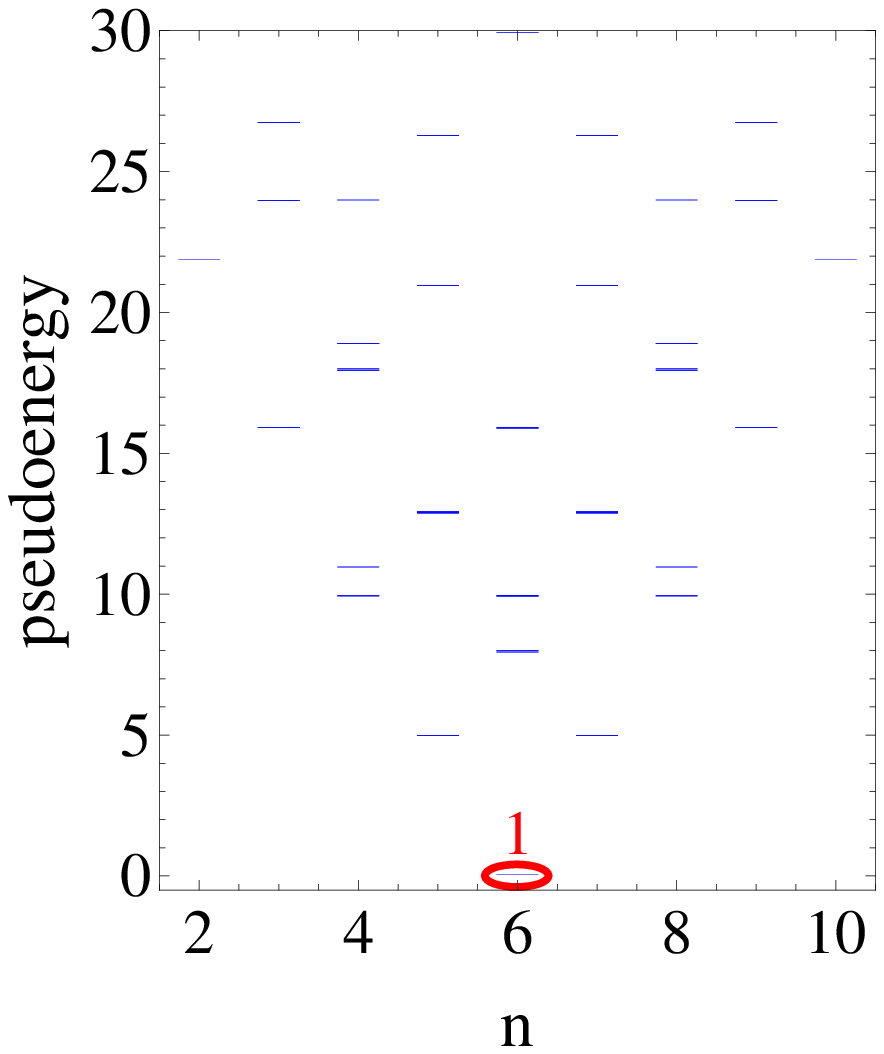}
\centering
\end{minipage}
\begin{minipage}{0.45\linewidth}
\small{(b)} $\delta t=-0.2,U=0$
\includegraphics[width=0.95\linewidth]{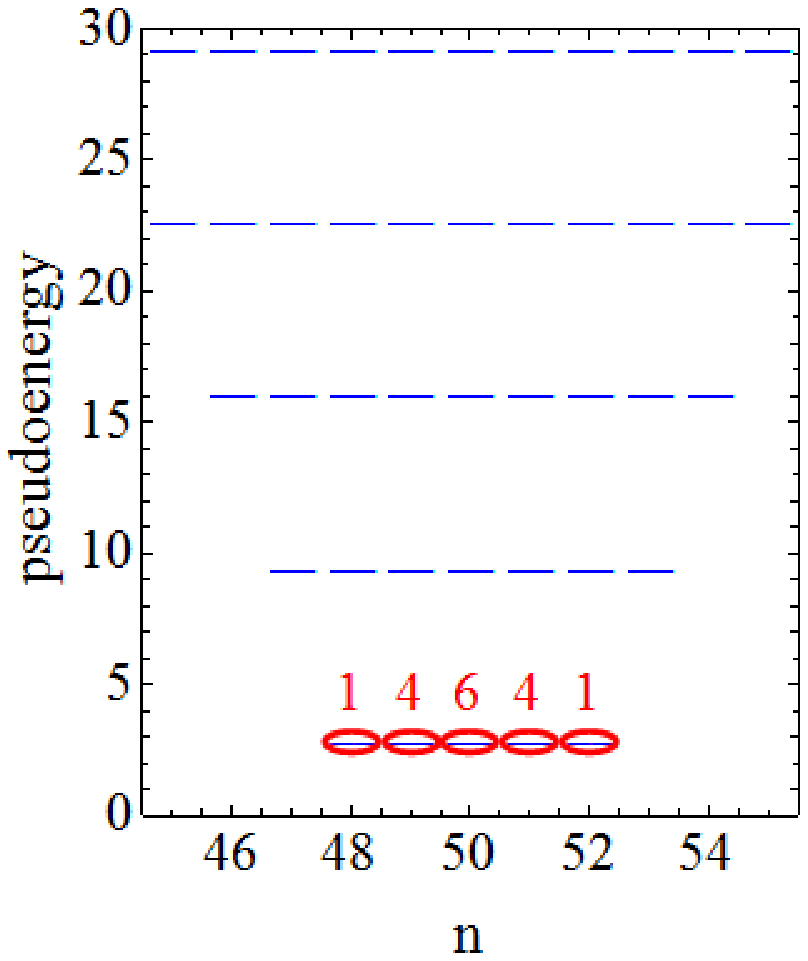}
\centering
\end{minipage}\\
\caption{\label{robustness} (a) The chain length $L$ is 12. We compare it with Fig.~\ref{interactingspectrum}. It shows that the degeneracy is robust when the size of the system changes. (b) The length L of the original chain is 200, and the subsystem is cut off with the length of 50 in both two figures. We compare it with Fig.~\ref{freespectrum}. It shows that the degeneracy is robust when the ratio of the subsystem in the whole system changes.}
\end{figure}

%\section{Robustness of the Method}
\emph{Robustness and physical interpretation.}---
Next, we consider the robustness of our results. As is shown in Figure.~\ref{robustness}, the meaning of robustness is in two aspects,
one is the length of the primary system and the other is the ratio of the subsystem in the whole system.
We just need to cut off an open chain (subsystem in OBC) from the primary closed chain (system in PBC), and the open chain does not have to be exactly a half of the primary closed chain. This is not surprising, since the method is based on the difference between OBC and PBC, which means that it is related with the topology and the change on the edge.

%\section{a Primary Explanation about the Validity of the Method}
%\emph{Interpretation of entanglement spectrum method for SSHH model.}---
We would like to point out that similar phenomena about the degeneracy of largest ES relating with physical properties are also referred in some other systems \cite{Pollmann,Rao}.
We now present a physical picture of how the method works by providing some detailed evidences.
We will still focus on SSHH model as an illustration. As we mentioned above, the method is related with bulk-edge correspondence. Considering two chains in PBC and OBC respectively, it is understandable that the primary parts of the state - the bulk parts - are very similar to each other
in the two boundary conditions.
When we cut off the open chain and obtain its reduced density matrix,
we can also obtain the eigenstates  in studying the ES, and they are certainly very similar with the basic
eigenstates of the OBC system. In topologically trivial phase, this leads to that there is only one leading eigenstate,
correspondingly the largest eigenvalue of the reduced density matrix in non-degenerate. However in the non-trivial phase, although the main components are similar with the bulk state in the open system, there is an uncertainty whether the edge mode is contained in the state. For example, if there is an edge mode in the open chain, the bulk state added with the edge mode or not is always the main components in the reduced density matrix, which means the low-lying ES will be two fold degenerate. The existence of the relation between edge state and ES in some free particle systems was studied in \cite{Sirker,Rao,Fidkowski}. Here, we give some more detailed evidences to show our explanations. In the non-trivial phase, there is four single-particle edge modes, so that we will have 16-fold degeneracy resulted from different combinations of whether each edge mode is contained. Also, we can count the number of remained particles corresponding to each degenerate state in theory in a simple way, and this is exactly the same with our numerical calculation.
\begin{figure}[h]
\begin{minipage}{0.49\linewidth}
\small{(a)}
\includegraphics[width=0.95\linewidth]{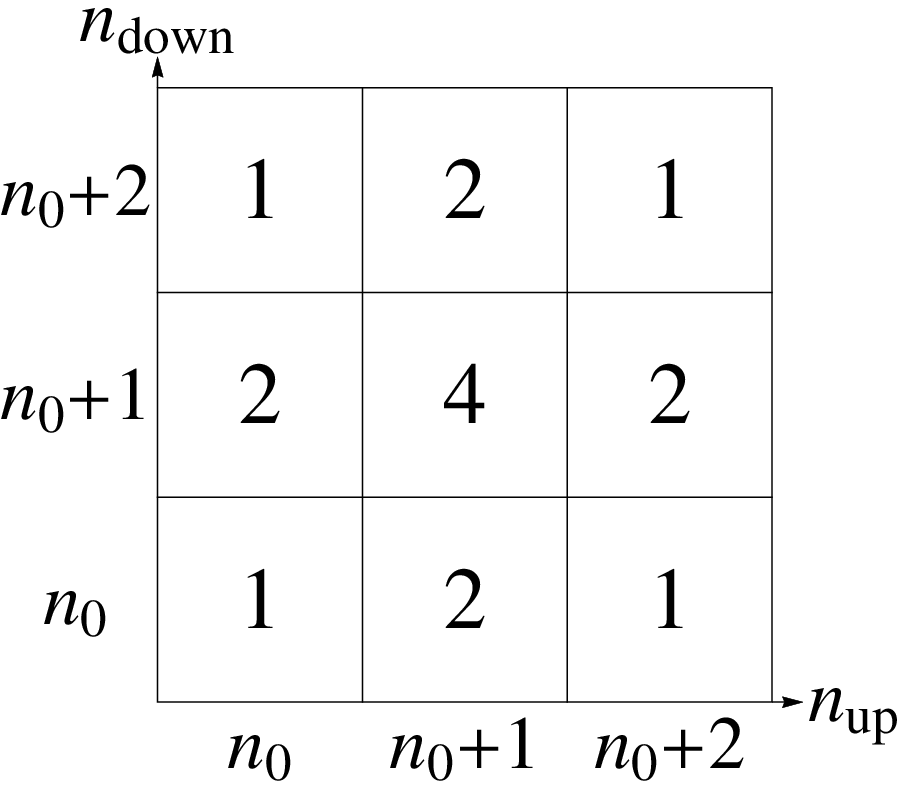}
\centering
\end{minipage}
\begin{minipage}{0.49\linewidth}
\small{(b)}
\includegraphics[width=0.95\linewidth]{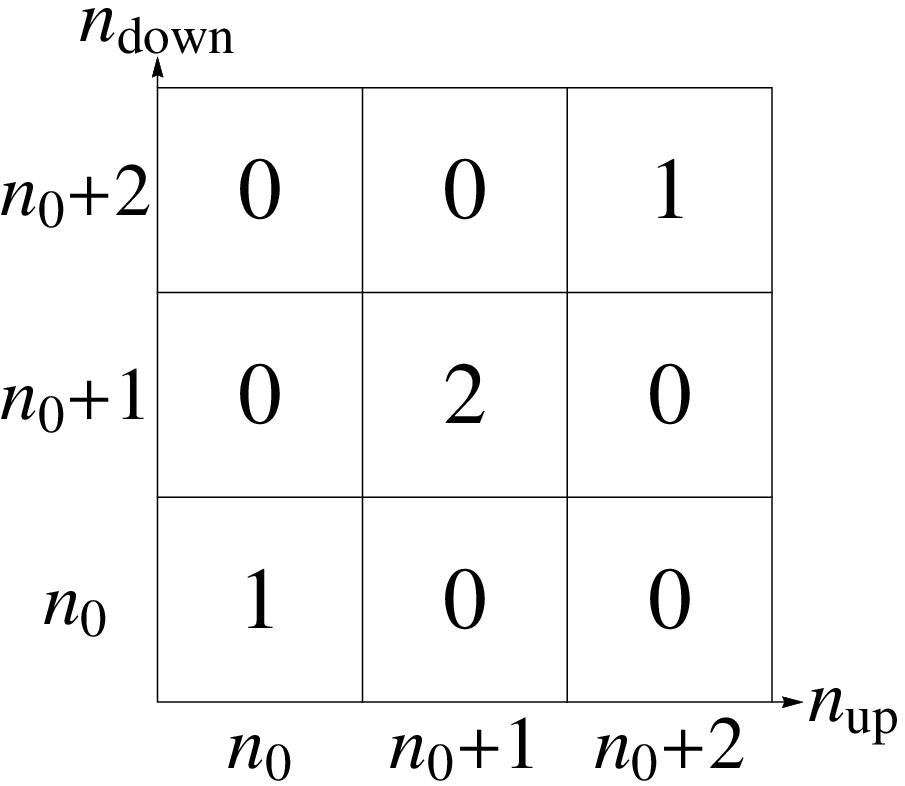}
\centering
\end{minipage}\\
\caption{\label{distribution} The numbers in the box show the distribution of degeneracy corresponding to different numbers of remaining particle with spin-up and spin-down in subsystem. The distributions are in (a)free fermion case: $U=0$, (b)interacting fermion case: $U<0$}
\end{figure}

This interpretation also works for case when the interaction is involved. Although the picture of single-particle edge state does not exist,
it is assumed that there is edge state of pseudo particle -- so called edge elementary excitation. The edge mode is often considered as a paired-particle mode when $U<0$, because  a pair of fermions each with spin-up and spin-down tend to be on the same site due to the attractive interaction.
The differences of particle number between the four degenerate spectral lines shows: a pair of particles with opposite spins are added or not into the bulk state of the whole system simultaneously when $U<0$. The result is shown in Fig. \ref{distribution}.
Also, this situation is similar for repulsive interaction $U>0$, since there is particle-hole symmetry in Hamiltonian of free fermions of each kind of spin in this half-filled system, we only need to see fermions with one kind of spin as holes so that the system is equivalent
with case of attractive interaction $U<0$. Here, we remark that the two different kinds of edge states
in attractive interaction $U<0$ and repulsive interaction $U>0$ correspond to different phases. This matches our phase diagram derived by investigating the distribution of degeneracy in ES. Our results also provide an evidence that the bulk-edge correspondence
exists for systems with interactions.

%\section{Conclusion}
\emph{Conclusion.}---
We distinguish different topological phases in SSHH model by dividing a closed chain into two open chains and detecting their ES. 
We only need to study the PBC case of the ground state, which is easier to solve than in OBC. Moreover, the lengths of both the closed chain and the ratio of cut-off open chain do not change the result. 
The bulk-edge correspondence and the change from PBC to OBC for entanglement 
confirm the validity of our method. To be specific, we give the evidence to show that the possible edge mode in PBC leads to 
the degeneracy of the ground state ES.
For the bulk-edge correspondence in interacting system, our explanation still applicable according to elementary excitation of edge mode,
for example we can consider whether the edge mode is single-particle, paired-particle or particle-hole-paired elementary excitation. 
Since our method is based on the difference between OBC and PBC (the topology of the system with only short-ranged interaction) in addition with the property on the edge, we expect that it is also valid in other similar systems.

\begin{acknowledgments}
This work is supported by NSFC (91536108), NFFTBS (J1030310, J1103205), the Strategic Priority Research Program of the Chinese Academy of Sciences (XDB01010000), the Young Elite Program for Faculty of Universities in Beijing, and Training Program of Innovation for Undergraduates of Beijing.
\end{acknowledgments}

\end{document}